\documentclass[prl,aps,amsmath,amssymb,reprint]{revtex4-1}
\usepackage{color}
\usepackage{amsmath}
\usepackage{amsthm}

\usepackage{bm}
\usepackage{multirow}
\usepackage{graphicx}
\usepackage{amsthm}
\begin{document}
	
\title{Desynchronization transitions in adaptive networks} %Title of paper

% repeat the \author .. \affiliation  etc. as needed
% \email, \thanks, \homepage, \altaffiliation all apply to the current author.
% Explanatory text should go in the []'s, 
% actual e-mail address or url should go in the {}'s for \email and \homepage.
% Please use the appropriate macro for the type of information

% \affiliation command applies to all authors since the last \affiliation command. 
% The \affiliation command should follow the other information.

\author{Rico Berner$^{1,2}$}
\email[]{rico.berner@physik.tu-berlin.de}
\author{Simon Vock$^{1}$}
\author{Eckehard Sch\"oll$^{1,3}$}
\author{Serhiy Yanchuk$^{2}$}
%\altaffiliation{}
\affiliation{$^{1}$Institut f\"ur Theoretische Physik, Technische Universit\"at Berlin, Hardenbergstr.\,36, 10623 Berlin, Germany}
\affiliation{$^{2}$Institut f\"ur Mathematik, Technische Universit\"at Berlin, Stra\ss e des 17. Juni 136, 10623 Berlin, Germany}
\affiliation{$^{3}$Bernstein Center for Computational Neuroscience Berlin, Humboldt-Universit\"at, Philippstra\ss e 13, 10115 Berlin, Germany}
% Collaboration name, if desired (requires use of superscriptaddress option in \documentclass). 
% \noaffiliation is required (may also be used with the \author command).
%\collaboration{}
%\noaffiliation

\date{\today}

\begin{abstract}
	Adaptive networks change their connectivity with time, depending on their dynamical state. While synchronization in structurally static networks has been studied extensively, this problem is much more challenging for adaptive networks.
	In this Letter, we develop the master stability approach for a large class of adaptive networks. This approach allows for reducing the synchronization problem for adaptive networks to a low-dimensional system, by decoupling topological and dynamical properties. 
	We show how the interplay between adaptivity and network structure gives rise to 
	the formation of stability islands. Moreover, we report a desynchronization transition and the emergence of complex partial synchronization patterns induced by an increasing overall coupling strength.
	We illustrate our findings using adaptive networks of coupled phase oscillators and FitzHugh-Nagumo neurons with synaptic plasticity. 
\end{abstract}

% insert suggested PACS numbers in braces on next line

\maketitle %\maketitle must follow title, authors, abstract and \pacs

%--------------------------------------------------
% Introduction
%--------------------------------------------------
%\section{Introduction}\label{sec:intro}
%\the\columnwidth
In nature and technology, complex networks serve as a ubiquitous paradigm with a broad range of applications from physics, chemistry, biology, neuroscience, socio-economic and other systems~\cite{NEW03}. Dynamical networks are composed of interacting dynamical units, such as, e.g., neurons or lasers. Collective behavior in dynamical networks has attracted much attention over the last decades. 
Depending on the network and the specific dynamical system, various synchronization patterns of increasing complexity were explored \cite{PIK01,STR01a,ARE08,BOC18}. Even in simple models of coupled oscillators, patterns such as complete synchronization~\cite{KUR84}, cluster synchronization~\cite{YAN01a,SOR07,BEL11a,GOL16,ZHA20}, and various forms of partial synchronization have been found, such as frequency clusters~\cite{BER19}, solitary~\cite{JAR18} or chimera states~\cite{KUR02a,ABR04,MOT10,PAN15,SCH16b,MAJ18a,OME19c,SCH19c,ZAK20}.
%The investigation of synchronization phenomena on complex networks~\cite{PIK01,STR01a,ARE08,BOC18} has become fruitful for applications in many areas ranging from physics and chemistry to biology, neuroscience, physiology, ecology, socio-economic systems, computer science and engineering. A well-known example is the synchronized flashing in groups of fireflies~\cite{BUC68}. 
In brain networks, particularly, synchronization is believed to play a crucial role: for instance, under normal conditions in the context of cognition and learning~\cite{SIN99,FEL11}, and under pathological conditions, such as Parkinson's disease~\cite{HAM07}, epilepsy~\cite{JIR13,JIR14,ROT14,AND16,GER20}, tinnitus~\cite{TAS12,TAS12a}, schizophrenia, to name a few~\cite{UHL09}. %With regards to pathological diseases, moreover, only recently the importance to study the mechanisms behind desynchronization were highlighted~\cite{TAN18}. 
Also in power grid networks, synchronization is essential for the stable operation~\cite{ROH12,MOT13a,MEN14,TAH19}. 
%The control of dynamics on networks is another important issue with a lot of applications~\cite{SCH07,SCH16}.

The powerful methodology of the master stability function~\cite{PEC98} has been a milestone for the analysis of synchronization phenomena. This method allows for separating dynamical from structural features for a given dynamical network. It drastically simplifies the problem by reducing the dimension and unifying the synchronization study for different networks. Since its introduction, the master stability approach has been extended and refined for multilayer~\cite{BRE18}, multiplex~\cite{TAN19,BER20} and hypernetworks~\cite{SOR12a,MUL20}; to account for single and distributed delays~\cite{CHO09,FLU10b,KEA12,KYR14,LEH15b,BOE20}; and to describe the stability of clustered states~\cite{DAH12,PEC14,GEN16,BLA19}. The master stability function has been used to understand effects in temporal~\cite{STI06,KOH14} as well as adaptive networks~\cite{ZHO06f} within a static formalism. Beyond the local stability described by the master stability function, Belykh et. al. have developed the connection graph stability method to provide analytic bounds for the global asymptotic stability of synchronized states~\cite{BEL04,BEL05b,BEL06,BEL06a}. Despite the apparent vivid interest in the stability features of synchronous states on complex networks, only little is known about the effects induced by an adaptive network structure. This lack of knowledge is even more surprising regarding how important adaptive networks are for the modeling of real-world systems.

Adaptive networks are commonly used models for synaptic plasticity~\cite{MAR97a,ABB00,CAP08a,MEI09a,MIK13,MIK14} which determines learning, memory, and development in neural circuits. Moreover, adaptive networks have been reported for chemical~\cite{JAI01,KUE19a}, epidemic~\cite{GRO06b}, biological~\cite{PRO05a}, transport~\cite{MAR17b}, and social systems~\cite{GRO08a,HOR20}. A paradigmatic example of adaptively coupled phase oscillators has recently attracted much attention~\cite{GUT11,ZHA15a,KAS17,ASL18a,KAS18,KAS18a,BER19,BER19a,BER20,FEK20}, and it appears to be useful for predicting and describing phenomena in more realistic and detailed models~\cite{POP15,LUE16,CHA17a,ROE19a}. Systems of phase oscillators are important for understanding synchronization phenomena in a wide range of applications~\cite{BRE10h,NAB11,BIC20}.

In this Letter, we report on a surprising desynchronization transition induced by an adaptive network structure. 
We find various parameter regimes of partial synchronization during the transition from the synchronized to an incoherent state. The partial synchronization phenomena include multi-frequency-cluster and chimera-like states. By going beyond the static network paradigm, we develop a master stability approach for networks with adaptive coupling. We show how the adaptivity of the network gives rise to the emergence of stability islands in the master stability function that result in the desynchronization transition. With this, we establish a general framework to study those transitions for a wide range of dynamical systems. In order to provide analytic insights, we use the generalized Kuramoto-Sakaguchi system on an adaptive and complex network. Finally, we show that our findings also hold for a more realistic neuronal set-up of coupled FitzHugh-Nagumo neurons with synaptic plasticity.

%--------------------------------------------------
% Model
%--------------------------------------------------
%\section{Model}
We consider the following general class of $N$ adaptively coupled systems  \cite{MAI07,GUT11,ZHA15a,KAS17,ASL18a,KAS18,KAS18a,BER19,BER19a,BER20}
\begin{align}
\dot{\bm{x}}_i&=f(\bm{x}_i)-{\sigma}\sum_{j=1}^N a_{ij}\kappa_{ij}g(\bm{x}_i,\bm{x}_j),\label{eq:adaptiveNW_x}\\
\dot{\kappa}_{ij} &= -\epsilon\left(\kappa_{ij} + a_{ij} h(\bm{x}_i-\bm{x}_j)\right), \label{eq:adaptiveNW_kappa}
\end{align}
where $\bm{x}_i\in \mathbb{R}^d$, $i=1,\dots,N$, is the $d$-dimensional dynamical variable of the $i$th node, $f(\bm{x}_i)$ describes the local dynamics of each node, and $g(\bm{x}_i,\bm{x}_j)$ is the coupling function. The coupling is weighted by scalar variables $\kappa_{ij}$ which are adapted dynamically according to Eq.~(\ref{eq:adaptiveNW_kappa}) with the nonlinear adaptation function $h(\bm{x}_i-\bm{x}_j)$. We assume that the adaptation depends on the difference of the corresponding dynamical variables, similar to the neuronal spike timing-dependent plasticity~\cite{ABB00,CAP08a,LAD13,COO16}. The base connectivity structure is given by the matrix elements ${a_{ij}\in\{0,1\}}$ of the $N\times N$ adjacency matrix $A$ which possesses a constant row sum $r$, i.e., $r=\sum_{j=1}^N a_{ij}$ for all $i=1,\dots,N$. The assumption of the constant row sum is necessary to allow for synchronization. The Laplacian matrix is $L=r\mathbb{I}_N -A$ where $\mathbb{I}_N$ is the $N$-dimensional identity matrix. The eigenvalues of $L$ are called Laplacian eigenvalues of the network. The parameter $\sigma >0$ defines the overall coupling input, and $\epsilon >0$ is a time-scale separation parameter. In particular, if the adaptation is slower than the local dynamics, the parameter $\epsilon$ is small.
% We note that the phase space of system ~\eqref{eq:adaptiveNW_x}--\eqref{eq:adaptiveNW_kappa} is $N(d+N)$ dimensional.

Complete synchronization is defined by the $N-1$ constraints $\bm{x}_1=\bm{x}_2=\cdots=\bm{x}_N$. Denoting the synchronization state by $\bm{x}_i(t)=\bm{s}(t)$ and $\kappa_{ij}=\kappa_{ij}^s$, we obtain from Eqs.~\eqref{eq:adaptiveNW_x}--\eqref{eq:adaptiveNW_kappa} the following equations for $\bm{s}(t)$ and $\kappa_{ij}^s$
\begin{align}
\dot{\bm{s}}&=f(\bm{s})+\sigma r h(0)g(\bm{s},\bm{s}),\label{eq:syncState_s}\\
{\kappa}^{s}_{ij} &= -a_{ij} h(0). \label{eq:syncState_kappa}
\end{align}
In particular, we see that $\bm{s}(t)$ satisfies the dynamical equation (\ref{eq:syncState_s}), and $\kappa_{ij}^s$ are either $-h(0)$ or zero, if the corresponding link in the base connectivity structure exists ($a_{ij}=1$) or not ($a_{ij}=0$), respectively. 

%--------------------------------------------------
% MSF theory
%--------------------------------------------------
%\section{MSF theory}
In order to describe the local stability of the synchronous state, we introduce the variations $\xi_i = \bm{x}_i-\bm{s}$ and ${\chi_{ij}}=\kappa_{ij}-{\kappa}^{s}_{ij}$. The linearized equations for these variations read
\begin{align}
\dot{\xi_i}&=\mathrm{D}f(\bm{s})\xi_i-{\sigma}g(\bm{s},\bm{s})\sum_{j=1}^N a_{ij}\chi_{ij}
\label{eq:var1}\\
&\quad+{\sigma}h(0)\sum_{j=1}^N a_{ij}(\mathrm{D}_1 g(\bm{s},\bm{s})\xi_i+\mathrm{D}_2 g(\bm{s},\bm{s})\xi_j),\nonumber \\
\dot{\chi}_{ij} &= -\epsilon\left(\chi_{ij} + a_{ij} \mathrm{D}h(0)
({\xi}_i-\xi_j)\right),\label{eq:var2}
\end{align}
where $\mathrm{D}f$ and $\mathrm{D}h$ are the Jacobians ($d \times d$ matrix and $1 \times d$ matrix, respectively), and $\mathrm{D}_1 g$ and $\mathrm{D}_2 g$ are the Jacobians with respect to the first and the second variable, respectively. 

The system (\ref{eq:var1})--(\ref{eq:var2}) is used to calculate the Lyapunov exponents of the synchronous state; it possesses very high dimension $N^2+Nd$. However, one can introduce a new coordinate frame which separates an $N(d+1)$-dimensional master from an $N(N-1)$-dimensional slave system. The new variables of the master system depend only on the variables of the master system itself and they are independent of the dynamics of the slave system. Further, we find that the dynamics of the slave system is ruled by the dynamics of the master system. With these new coordinates, we reduce the system's dimension significantly. Moreover, as in the classical master stability approach, we diagonalize the $N(d+1)$-dimensional master system into blocks of $d+1$ dimensions. Hence, the dynamics in each block is described by the new coordinates $\zeta$ and $\kappa$ which are $d$- and one-dimensional dynamical variables, respectively. Our analysis shows that the coupling structure enters just as a complex parameter $\mu$, the network's Laplacian eigenvalue. For all details and the proof of the master stability function, we refer to the Supplemental Material~\cite{suppl}.

As a result, the stability problem is reduced to the largest Lyapunov exponent $\Lambda(\mu)$, depending on a complex parameter $\mu$, for the following system 
\begin{align}
\begin{split}
\dot \zeta &= \bigg(\mathrm{D}f(\bm{s})+\sigma r h(0)\big[\mathrm{D_1}g(\bm{s},\bm{s}) \\
&\quad\quad + (1-\frac{\mu}{r}) \mathrm{D_2}g(\bm{s},\bm{s})\big]\bigg) \zeta - {\sigma} g(\bm{s},\bm{s}) \kappa, 
\end{split} \label{eq:adaptiveNW_MSF_zeta}\\ 
\dot \kappa &=  -\epsilon\left(\mu \mathrm{D}h(0) \zeta + \kappa\right),\label{eq:adaptiveNW_MSF_kappa}.
\end{align}
The function $\Lambda(\mu)$ is called master stability function.  Note that the first bracketed term in $\zeta$ of~\eqref{eq:adaptiveNW_MSF_zeta} resembles the master stability approach for static networks, which, in this case, is equipped by an additional interaction representing the adaptation.

%--------------------------------------------------
% Stability island induced by adaptivity
%--------------------------------------------------
%\section{Stability island induced by adaptivity}
To obtain analytic insights into the stability features of synchronous states that are induced by an adaptive coupling structure, we consider the following model of $N$ adaptively coupled phase oscillators~\cite{KAS17,BER19}
\begin{align}
\dot{\phi}_i &= \omega + \sigma \sum_{j=1}^N a_{ij}\kappa_{ij} \sin(\phi_i-\phi_j+\alpha), \label{eq:APO_phi}\\
\dot{\kappa}_{ij} & = -\epsilon\left(\kappa_{ij} + a_{ij} \sin(\phi_i - \phi_j + \beta)\right), \label{eq:APO_kappa}
\end{align}
where $\phi_{i}$ represents the phase of the $i$th oscillator, $\omega$ is its natural frequency which we set to zero in a rotating frame. 

The synchronous state of~\eqref{eq:APO_phi}--\eqref{eq:APO_kappa} is given by $\bm{s}(t)=(\sigma r \sin\alpha\sin\beta) t$ and $\kappa_{ij}^s=-a_{ij}\sin\beta$. Using~\eqref{eq:adaptiveNW_MSF_zeta}--\eqref{eq:adaptiveNW_MSF_kappa}, the stability of the synchronous state is described by the quadratic characteristic polynomial
\begin{align}\label{eq:MSF_Kuramoto_eigenvalues}
\lambda^2 + \left(\epsilon-\sigma{\mu}\cos(\alpha)\sin(\beta)\right)\lambda -\epsilon\sigma{\mu}\sin(\alpha+\beta) = 0.
\end{align}
The master stability function for the synchronous solution is given as the maximum real part $\Lambda=\max \mathrm{Re} (\lambda_{1,2})$ of the solutions $\lambda_{1,2}$ of the polynomial~\eqref{eq:MSF_Kuramoto_eigenvalues}. These solutions $\lambda_{1,2}$ should be considered as functions of the complex parameter $\mu$ determining the network structure. It is convenient, however, to use the parameter $\sigma\mu$ in our case.

\begin{figure}
	\includegraphics{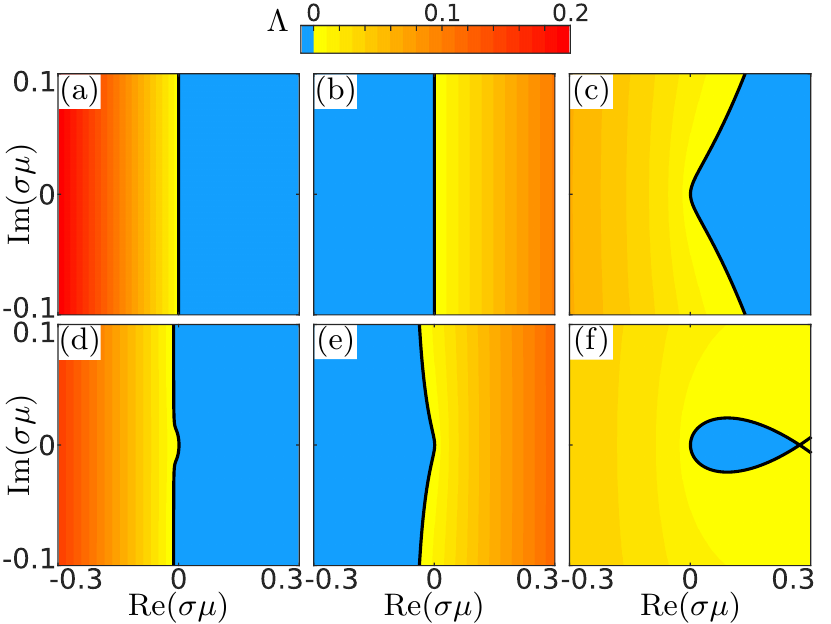}
	\caption{Master stability function $\Lambda(\mu)$ for the adaptive phase oscillator network \eqref{eq:APO_phi}--\eqref{eq:APO_kappa}. Regions belonging to negative Lyapunov exponents  $\Lambda$ are colored blue. The curve where $\Lambda(\mu)=0$ is given as a black solid line. In panels (a) and (b) the case without adaptation ($\epsilon=0$) is presented for $\beta=-0.35\pi$ and $\beta=0.2\pi$, respectively. Other panels: $\epsilon=0.01$ and (c) $\beta=-0.95\pi$, (d) $\beta=-0.35\pi$, (e) $\beta=0.2\pi$, and (f) $\beta=0.98\pi$. In all panels $\alpha=0.3\pi$. \label{fig:MSF_Kuramoto}}
\end{figure}
Figure~\ref{fig:MSF_Kuramoto} displays the master stability function determined for different adaptation rules controlled by $\beta$. The blue-colored areas correspond to regions that lead to stable dynamics. By changing the control parameter $\beta$, various shapes of the stable regions are visible. For some parameters, e.g., Fig.~\ref{fig:MSF_Kuramoto}(c,d,e), almost a whole half-space either left or right of the imaginary axis belongs to the stable regime. This resembles the case of no adaptation where the stability of the synchronous state is solely described by the sign of the real part of $\sigma\mu\sin\beta\cos\alpha$, see Fig.~\ref{fig:MSF_Kuramoto}(a,b). We also find parameters where most values $\sigma \mu$ correspond to unstable dynamics, except for an island, i.e., a bounded region in $\sigma \mu$ parameter space, see Fig.~\ref{fig:MSF_Kuramoto}(f).

To understand the emergence of the stability islands, we analyze the boundary that separates the stable ($\Lambda<0$) from the unstable region ($\Lambda>0$). This boundary is given by the condition $\Lambda=\mathrm{Re}\lambda=0$, or, equivalently, $\lambda=\mathrm{i}\gamma$. Substituting this into  Eq.~\eqref{eq:MSF_Kuramoto_eigenvalues}, we obtain a parameterized expression for the boundary as a function of $\gamma$ that has the form $\sigma\mu=Z(\gamma)$ with $Z(\gamma)$ given explicitly in the Supplemental material~\cite{suppl}. The latter parametrization of the boundary is displayed in Fig.~\ref{fig:MSF_Kuramoto} as the solid black line. It is straightforward to show that a stability island exists if $\sin(\alpha+\beta)/(\cos\alpha\sin\beta)<0$. The latter condition indicates a certain balance between the coupling and adaptation function. We emphasize that the emergence of stability islands is a direct consequence of adaptation. Without adaptation, the boundary simplifies to the axis $\mathrm{Re} \,\mu=0$, see Figs.~\ref{fig:MSF_Kuramoto}(a,b).

%--------------------------------------------------
% Desynchronization transition with increasing overall coupling 
%--------------------------------------------------
%\section{Desynchronization transition with increasing overall coupling}
In the following, we analyze the behavior of the adaptive network of phase oscillators \eqref{eq:APO_phi}--\eqref{eq:APO_kappa} in the presence of a stability island, and show how such an island introduces a desynchronization transition with increasing overall coupling $\sigma$. To measure the coherence, we use the cluster parameter $R_C$~\cite{KAS17,KAS18a}, which is given by the number of pairwise coherent oscillators normalized by the total number of pairs $N^2$. In the case of complete synchronization, frequency clustering, or incoherence, the cluster parameter values are $R_C=1$, $1<R_C<0$, or $R_C=0$, respectively, see Supplemental Material for details~\cite{suppl}.

\begin{figure}
	\includegraphics{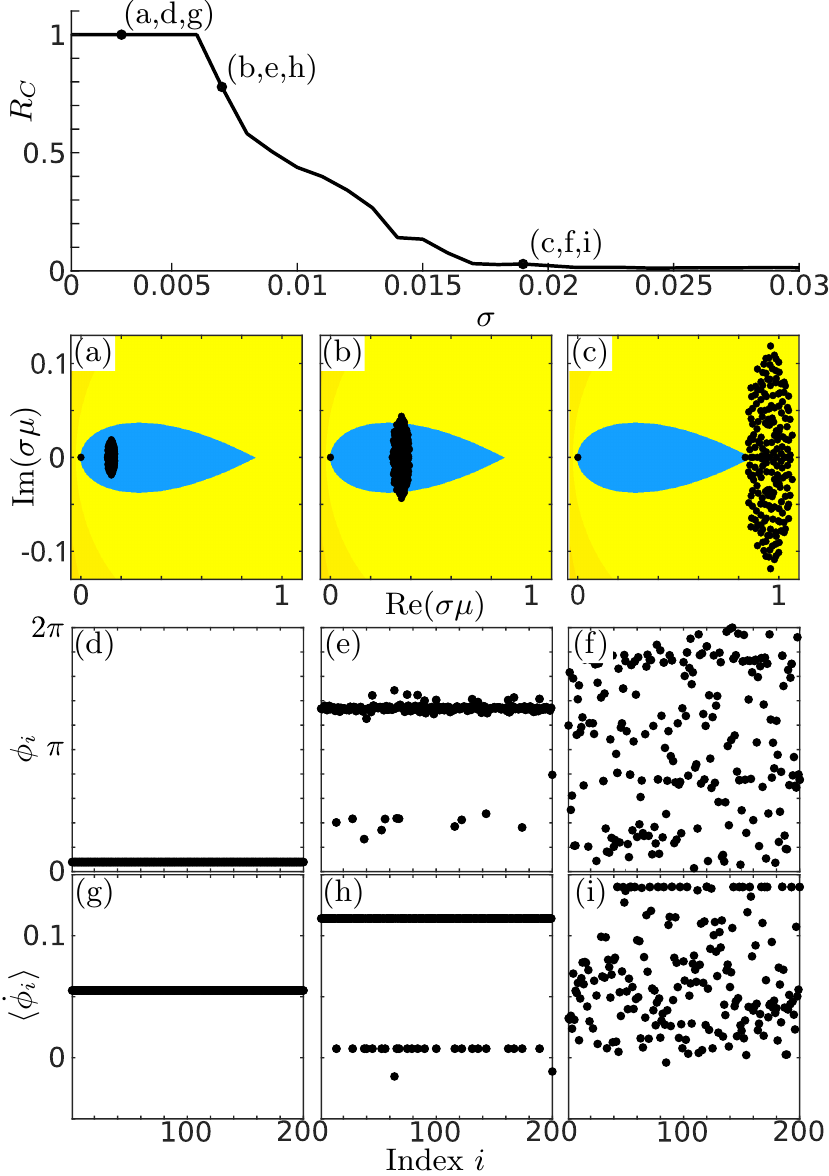}
	\caption{ Dynamics in the network of $200$ oscillators~\eqref{eq:APO_phi}--\eqref{eq:APO_kappa} with random adjacency matrix $A_\text{c}$~\cite{suppl}, and different values of overall coupling strength $\sigma$. Adiabatic continuation for increasing $\sigma$ with the stepsize of $0.001$, starting with the synchronous state $\phi_i=0$, $\kappa_{ij}=-a_{ij}\sin\beta$. The top panel shows the cluster parameter $R_C$ vs $\sigma$. For the three values of $\sigma$: (a,d,g) $\sigma=0.003$, (b,e,h) $\sigma=0.007$, and (c,f,i) $\sigma=0.019$,	the plots show: in (a,b,c) the master stability function color coded as in Fig. 1, together with $\sigma\mu_i$, where $\mu_i$ are the $N$ Laplacian eigenvalues of $A_c$; in (d,e,f) snapshots for $\phi_i$ at $t=30000$; and in (g,h,i) the temporal average of the phase velocities $\langle\dot{\phi}_i\rangle$ over the last $5000$ time units. Other parameters: $\alpha=0.49\pi$, $\beta=0.88\pi$, $\epsilon=0.01$. \label{fig:MSF_complex_MC}}
\end{figure}
The top panel in Fig.~\ref{fig:MSF_complex_MC} shows the cluster parameter $R_C$ for different values of the overall coupling constant $\sigma$. We observe that for small $\sigma$, the synchronous state is stable, see Fig.~\ref{fig:MSF_complex_MC}(a,d,g). This stability follows directly from the master stability function since all values $\sigma\mu_i$ for all Laplacian eigenvalues lie within the stability island, see Fig.~\ref{fig:MSF_complex_MC}(a).

By increasing the coupling strength $\sigma$, the values $\sigma\mu_i$ move out of the stability island ($\mu_i$ remain the same), and the synchronous state becomes unstable, see Fig.~\ref{fig:MSF_complex_MC}(b,c). For intermediate values of $\sigma$, multiclusters with hierarchical structure in the cluster size emerge, see Fig.~\ref{fig:MSF_complex_MC}(e,h) for a three-cluster state. Increasing the coupling constant further leads to the emergence of incoherence. In Fig.~\ref{fig:MSF_complex_MC}(f,i), the coexistence of a coherent and an incoherent cluster is presented. Such chimera-like states have been numerically studied for adaptive networks in \cite{KAS17,KAS18,KAS18a}. %We point out that for the cases in Figs.~\ref{fig:MSF_complex_MC}(e,h) and \ref{fig:MSF_complex_MC}(f,i) even the stability of the single coherent clusters can be deduced from the master stability function effectively, see~\cite{suppl} for details.

%--------------------------------------------------
% Application to neuronal network with synaptic plasticity
%--------------------------------------------------
%\section{Application to neuronal network with synaptic plasticity}
In the following, we show how our findings are transferred to a more realistic set-up of coupled neurons with synaptic plasticity. For this, we consider a network of FitzHugh-Nagumo neurons~\cite{STE08,OME13,GER14a,BAS18} coupled through chemical excitatory synapses~\cite{DRO04,WEC05,LI07b} equipped with plasticity:
\begin{align}
\tau \dot u_i &= u_i - \frac{u_i^3}{3} - v_i - {\sigma}u_i\sum_{j = 1}^{N} a_{ij} 
\kappa_{ij} { I_j},\label{eq:AFHN_FHN1}\\
\dot v_i &= u_i + a - b v_i,\label{eq:AFHN_FHN2}\\
\dot I_i &=  \alpha(u_i) (1-I_i)- {I_i}/{\tau_\text{syn}},\label{eq:AFHN_FHN3}\\
\dot{\kappa}_{ij} &= -\epsilon \left( \kappa_{ij} + a_{ij} e^{-\beta_1 (u_i - u_j + \beta_2)^2} \right).\label{eq:AFHN_kappa}
\end{align}
Here $u_i$ denotes the membrane potential and $v_i$ summarizes the recovery processes for each neuron; $I_i$ describes the synaptic output for each neuron; the parameters $a=0.7$ and $b=0.2$ are fixed to the values corresponding to self-sustained oscillatory dynamics of uncoupled neurons; and $\tau=0.08$ and $\epsilon=0.01$ are fixed time scale separation parameters between the fast activation and slow inhibitory processes in each neuron, and between the fast oscillatory dynamics and the slow adaptation of the coupling weights, respectively. The synaptic recovery function is given by $\alpha(u) = 2/(0.08(1 + \exp(-{u}/0.05)))$. The synaptic timescale is $\tau_\text{syn}=5/6$. For more details on the model, we refer to~\cite{LI07b,suppl}. The form of the synaptic plasticity is similar to the rules used in~\cite{YUA11,CHA17a}. We consider $\beta_1$ and $\beta_2$ as control parameters of the adaptation function. Note that $\beta_1$ and $\beta_2$ are uniquely determined by the values of $h(0)$ and $\mathrm{D}h(0)$ of the plasticity rule, %in Eq.~\eqref{eq:AFHN_kappa}
and these are the only essential parameters of the plasticity function, regarding the stability of the synchronous state, see Eqs.~\eqref{eq:adaptiveNW_MSF_zeta}--\eqref{eq:adaptiveNW_MSF_kappa}. 

The synchronous state of the network of FitzHugh-Nagumo neurons \eqref{eq:AFHN_FHN1}--\eqref{eq:AFHN_kappa} satisfies Eqs.~\eqref{eq:syncState_s}--\eqref{eq:syncState_kappa}, and it is periodic for the chosen parameter values. Using our extended master stability approach, we determine numerically the master stability function which is the maximum Lyapunov exponent of Eqs.~\eqref{eq:adaptiveNW_MSF_zeta}--\eqref{eq:adaptiveNW_MSF_kappa}. 

\begin{figure}
	\includegraphics{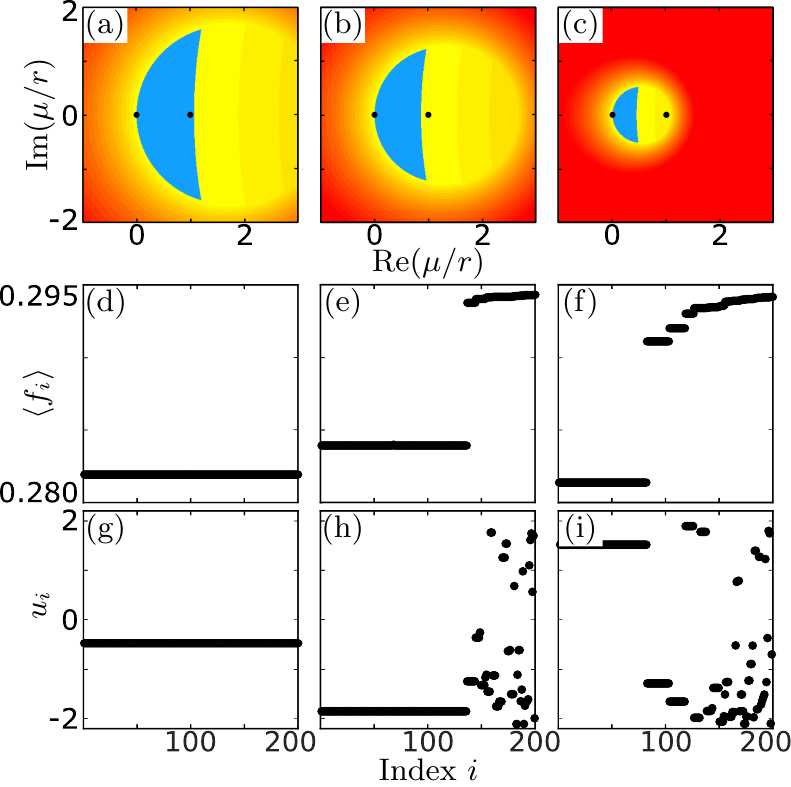}
	\caption{Dynamics of globally coupled network of $200$ FitzHugh-Nagumo neurons with plasticity Eqs.~\eqref{eq:AFHN_FHN1}--\eqref{eq:AFHN_kappa}. Adiabatic continuation for an increasing overall coupling strength $\sigma$ with the step size $0.0005$, starting with the synchronous state. For the three values of $\sigma$: (a,d,g) $\sigma=0.002$, (b,e,h) $\sigma=0.0025$, and (c,f,i) $\sigma=0.005$, the plots show: in (a,b,c), the master stability function, together with $\mu_i/r$, where $\mu_i$ are the Laplacian eigenvalues (color code as in Fig.~\ref{fig:MSF_Kuramoto}), in (d,e,f) the average frequency $\langle f_i\rangle$, and in (g,h,i) snapshots for $u_i$ at $t=10000$. Here $\langle f_i\rangle=M_i/1000$, where  $M_i$ is the number of rotations (spikes) of neuron $i$ during the time interval of length 1000. The control parameters for the adaptation rule $\beta_1$ and $\beta_2$ are chosen such that $h(0)=0.8$ and $\mathrm{D}h(0)=(80,0,0)$.
		\label{fig:MSF_global_FHN}}
\end{figure}

In Fig.~\ref{fig:MSF_global_FHN}(a,b,c), we show the master stability function in dependence on the parameter $\mu/r$ for different values of the overall coupling constant $\sigma$. We observe a stability island for the chosen set of parameters, see the Supplemental material for other parameter values~\cite{suppl}. In contrast to the phase oscillator network in Fig.~\ref{fig:MSF_complex_MC}, the master stability function does not scale linearly with $\sigma$. This is due to the non-diffusive coupling function in Eq.~\eqref{eq:AFHN_FHN1}. Moreover, with increasing $\sigma$, the size of the stability island shrinks. Since all Laplacian eigenvalues $\mu_i$  are independent of $\sigma$, we observe that $\mu_i/r$ move out of the stability island with increasing $\sigma$. For the globally coupled network, in particular, we have either $\mu_i/r=0$ or $\mu_i/r=1$. Therefore, with increasing $\sigma$, we find a transition from complete coherence, see Fig.~\ref{fig:MSF_global_FHN}(a,d,g) to partial synchronization and incoherence. We further observe that closely after destabilization, a large frequency cluster remains visible, see Fig.~\ref{fig:MSF_global_FHN}(b,e,h). For higher overall coupling, the cluster sizes shrink, and the number of small clusters increases, see Fig.~\ref{fig:MSF_global_FHN}(c,f,i).

%--------------------------------------------------
% Conclusions
%--------------------------------------------------
%\section{Conclusions}
In summary, we have developed a master stability approach for a general class of adaptive networks. This approach allows for studying the subtle interplay between nodal dynamics, adaptivity, and a complex network structure. The master stability approach has been first applied to a paradigmatic model of adaptively coupled phase oscillators. We have presented several typical forms of the master stability function for different adaptation rules, and observed adaptivity-induced stability islands. Besides, we have shown that stability islands give rise to the emergence of multicluster states and chimera-like states in the desynchronization transition for an increasing overall coupling strength. Qualitatively the same phenomena have been shown for a more realistic network of non-diffusively coupled FitzHugh-Nagumo neurons with synaptic plasticity. In this set-up, the emergence of a stability island and a desynchronization transition have been found as well.

The theoretical approach introduced in this Letter provides a powerful tool to study collective effects in more realistic neuronal network models, including synaptic plasticity~\cite{TAS12a,POP15}. While our approach is presented for differentiable models, it might be generalized to non-continuous models of spiking neurons equipped with spike timing-dependent plasticity~\cite{LAD13,COO16}. %Moreover, it has been shown that an adaptive network structure features multistability and the emergence of divers partial synchronization patterns~\cite{MAI07,KAS17,BER19}. 
Our findings on the transition from coherence to incoherence reveal the role adaptivity plays for the formation of partially synchronized patterns which are important for understanding the functioning of neuronal systems~\cite{TAN18}. Beyond neuronal networks, adaptation is a well-known control paradigm~\cite{LEL10,YU12,LEH14,SCH16}. Our extended master stability approach provides a generalized framework to study various adaptive control schemes for a wide range of dynamical systems.

\begin{acknowledgments}
	This work was supported by the German Research Foundation DFG, Project Nos. 411803875 and 440145547. 
\end{acknowledgments}

\clearpage
%\thispagestyle{empty}
%\addtocounter{page}{-1}
%\title{}

\onecolumngrid
\begin{center}
	\textbf{\large Supplemental Material on:\\ Desynchronization transitions in adaptive networks}\\[.2cm]
	Rico Berner$^{1,2,*}$, Simon Vock$^{1}$, Eckehard Sch\"oll$^{1,3}$, and Serhiy Yanchuk$^{2}$\\[.1cm]
	{\itshape $^{1}$Institut f\"ur Theoretische Physik, Technische Universit\"at Berlin, Hardenbergstr. 36, 10623 Berlin, Germany\\
		$^{2}$Institut f\"ur Mathematik, Technische Universit\"at Berlin, Stra\ss e des 17. Juni 136, 10623 Berlin, Germany\\
		$^{3}$Bernstein Center for Computational Neuroscience Berlin, Humboldt-Universit\"at, Philippstra\ss e 13, 10115 Berlin, Germany\\}
	%(Dated: \today)\\[1cm]
\end{center}
\twocolumngrid
\setcounter{equation}{0}
\setcounter{figure}{0}
\setcounter{table}{0}
\setcounter{page}{1}
\renewcommand{\theequation}{S\arabic{equation}}
\renewcommand{\thefigure}{S\arabic{figure}}
\renewcommand{\thetable}{S\arabic{table}}
\renewcommand{\bibnumfmt}[1]{[S#1]}
\renewcommand{\citenumfont}[1]{#1}
\setcounter{secnumdepth}{1}

%--------------------------------------------------
% The master stability approach for adaptive complex networks with non-diffusive coupling
%--------------------------------------------------
\section{Derivation of the master stability function for adaptive complex networks}\label{sec:Aug_MSA_MltPlx}
In this section, we derive the master stability function for system~\eqref{eq:adaptiveNW_x}--\eqref{eq:adaptiveNW_kappa} from the main text. For convenience, we repeat these equations here:
\begin{align}
\dot{\bm{x}}_i&=f(\bm{x}_i)-{\sigma}\sum_{j=1}^N a_{ij}\kappa_{ij}g(\bm{x}_i,\bm{x}_j),\label{eq:suppl_adaptiveNW_x}\\
\dot{\kappa}_{ij} &= -\epsilon\left(\kappa_{ij} + a_{ij} h(\bm{x}_i-\bm{x}_j)\right), \label{eq:suppl_adaptiveNW_kappa}
\end{align}
where the adjacency matrix has constant row sum  $r=\sum_{j=1}^N a_{ij}$.

Let $(\bm{s}(t),\kappa^s_{ij})$ be the synchronous state, i.e., $\bm{x}_i=\bm{s}(t)$ and $\kappa_{ij}=\kappa^s_{ij}$ for all $i,j=1\dots,N$. This state solves the set of differential Eqs.~\eqref{eq:syncState_s}--\eqref{eq:syncState_kappa} of the main text.

In order to describe the local stability of the synchronous state, we derive the variational equation for small perturbations close to this state. For this, we introduce the following vector variables denoting the deviations from the synchronized state:
$\bm{\xi}=\bm{x}-\mathbb{I}_N\otimes\bm{s}$, and $\bm{\chi}=\bm{\kappa}-\bm{\kappa}^{s}$ with
\begin{align*}
\bm{x} &= (\bm{x}_1,\cdots,\bm{x}_N)^T,\\
\bm{\kappa} &= (\kappa_{11},\cdots,\kappa_{1N},{\kappa}_{21},\cdots,\kappa_{NN})^T,
\end{align*}
where $\otimes$ denotes the Kronecker product. Using the following notations
\begin{align*}
\mathbf{a}_i&=(a_{i1},\dots,a_{iN}),\\
\mathrm{diag}(\mathbf{a}_i) &= \begin{pmatrix}
a_{i1} & & \\
& \ddots & \\
& & a_{iN}
\end{pmatrix},
\end{align*}
and the $N\times N^2$, $N^2\times N$, and $N^2\times N$ matrices
\begin{align*}
B&= \begin{pmatrix}
\mathbf{a}_{1} & & \\
& \ddots & \\
& & \mathbf{a}_{N}
\end{pmatrix}, \\
C&=B^T-{D}, \\
D &= \begin{pmatrix}
\mathrm{diag}(\mathbf{a}_1) \\
\vdots \\
\mathrm{diag}(\mathbf{a}_N)
\end{pmatrix},
\end{align*}
respectively, the variational equation reads 
\begin{align}\label{eq:VarEqs}
\begin{pmatrix}
\dot{\bm{\xi}} \\
\dot{\bm{\chi}}
\end{pmatrix}
=\begin{pmatrix}
S & -{\sigma} B\otimes g(\bm{s},\bm{s}) \\ 
-\epsilon C\otimes \mathrm{D}h(0) & -\epsilon\mathbb{I}_{N^2}
\end{pmatrix}
\begin{pmatrix}
{\bm{\xi}} \\
{\bm{\chi}}
\end{pmatrix},
\end{align}
where 
\begin{multline*}
S=\mathbb{I}_{N}\otimes \mathrm{D}f(\bm{s})\\+{\sigma} h(0)\left(r\mathbb{I}_N\otimes\mathrm{D_1}g(\bm{s},\bm{s}) + A\otimes \mathrm{D_2}g(\bm{s},\bm{s})\right).
\end{multline*}

We note that matrices $B,C$, and $D$ satisfy the relations $B\cdot B^T = r\mathbb{I}_N$, $B\cdot D = A$,  and $B\cdot C = L$, which can be obtained by straightforward calculation.

Due to the structure of the variational equation~\eqref{eq:VarEqs}, there exist $N^2-N$ eigenvalues $\lambda=-\epsilon$. 
The corresponding time-independent eigenspace can be found from
\begin{align*}
\begin{pmatrix}
S -\epsilon\mathbb{I}_{Nd} & -{\sigma} B\otimes g(\bm{s},\bm{s}) \\ 
-\epsilon C\otimes \mathrm{D}h(0) & 0
\end{pmatrix}
\begin{pmatrix}
{\bm{\xi}} \\
{\bm{\chi}}
\end{pmatrix} = 0.
\end{align*}
One can see that $(\bm{\xi},\bm{\chi})$ such that $\bm{\xi}=0$ and $B\bm{\chi}=0$ are the time-independent eigenvectors. Moreover, the relation $B\bm{\chi}=0$ defines $N^2-N$ linearly independent eigenvectors spanning the eigenspace corresponding to the eigenvalues $\lambda=-\epsilon$. This follows from the fact that $\bm{\chi}$ is $N^2$-dimensional and $\mathrm{rank}(B)=N$ if the row sum $r$ of $A$ is non-zero.

With these prerequisites we are now able to simplify the local stability analysis on adaptive networks and find a master stability function.
\\
%\label{thm:MSF_AdaptiveCmplSyst}
\textit{Let \eqref{eq:suppl_adaptiveNW_x}--\eqref{eq:suppl_adaptiveNW_kappa} possess a synchronous solution $(\bm{s},\kappa^s_{ij})$. Further, let \eqref{eq:VarEqs} be the variational equations around this synchronous solution and assume that the Laplacian matrix $L$ is diagonalizable. Then, the synchronous solution is locally stable if and only if for all eigenvalues $\mu\in\mathbb{C}$ of the Laplacian matrix, the largest Lyapunov exponent (if it exists), i.e., the master stability function $\Lambda(\mu)$, of the following system is negative}
\begin{align}
\begin{split}
\frac{\mathrm{d}\zeta}{\mathrm{d}t} &= \bigg(\mathrm{D}f(\bm{s})+\sigma r h(0)\big(\mathrm{D_1}g(\bm{s},\bm{s}) \\
&\quad\quad + (1-\frac{\mu}{r}) \mathrm{D_2}g(\bm{s},\bm{s})\big)\bigg) \zeta - {\sigma} g(\bm{s},\bm{s}) \kappa, 
\end{split} \label{eq:suppl_adaptiveNW_MSF_zeta}\\ 
\frac{\mathrm{d}\kappa}{\mathrm{d}t} &=  -\epsilon\left(\mu \mathrm{D}h(0) \zeta + \kappa\right).\label{eq:suppl_adaptiveNW_MSF_kappa}
\end{align}
Here, $\zeta\in\mathbb{C}^d$ and $\kappa\in\mathbb{C}$.

In the following we present the derivation of \eqref{eq:suppl_adaptiveNW_MSF_zeta}--\eqref{eq:suppl_adaptiveNW_MSF_kappa}.	As it is shown above, there are $N^2-N$ independent vectors $\bm{w}_l$ ($l=1,\dots,N^2-N$) spanning the kernel of $B$, i.e. $B \bm{w}_l =0$. Using the Gram-Schmidt procedure we find an orthonormal basis for $\ker(B)=\mathrm{span}\{\bm{v}_1,\dots,\bm{v}_{N^2-N}\}$. With this, we define the $N^2\times (N^2-N)$ matrix $Q=\left(\bm{v}_1,\dots,\bm{v}_{N^2-N}\right)$. Consider now the $(N^2+Nd)\times (N^2+Nd)$ matrix 
\begin{align*}
R = \begin{pmatrix}
\mathbb{I}_{Nd} & 0 & 0\\
0 & (1/r)B^T & Q
\end{pmatrix}
\end{align*}
with left inverse
\begin{align*}
R^{-1} = \begin{pmatrix}
\mathbb{I}_{Nd} & 0\\
0 & B \\
0 & Q^T
\end{pmatrix},
\end{align*}
i.e., $R^{-1} R = \mathbb{I}_{N^2+Nd}$. Introduce the new coordinates given by $R\begin{pmatrix}
\hat{{\bm{\xi}}} \\
\hat{{\bm{\chi}}}
\end{pmatrix}=\begin{pmatrix}
{\bm{\xi}} \\
{\bm{\chi}}
\end{pmatrix}$ for which the variational equation then reads
\begin{align*}
\frac{\mathrm{d}}{\mathrm{d}t}
\begin{pmatrix}
\hat{{\bm{\xi}}} \\
\hat{{\bm{\chi}}}
\end{pmatrix}
=
R^{-1}
\begin{pmatrix}
S & -{\sigma} B\otimes g(\bm{s},\bm{s}) \\ 
-\epsilon C\otimes \mathrm{D}h(0) & -\epsilon\mathbb{I}_{N^2}
\end{pmatrix}
R
\begin{pmatrix}
\hat{{\bm{\xi}}} \\
\hat{{\bm{\chi}}}
\end{pmatrix}.
\end{align*}
We further obtain
\begin{multline*}
R^{-1}
\begin{pmatrix}
S & -{\sigma} B\otimes g(\bm{s},\bm{s}) \\ 
-\epsilon C\otimes \mathrm{D}h(0) & -\epsilon\mathbb{I}_{N^2}
\end{pmatrix}
R\\
=R^{-1}
\begin{pmatrix}
S & -{\sigma} \mathbb{I}_N\otimes g(\bm{s},\bm{s}) & 0 \\ 
-\epsilon C\otimes \mathrm{D}h(0) & -\epsilon/r B^T & -\epsilon Q
\end{pmatrix}\\
=\begin{pmatrix}
S & -{\sigma} \mathbb{I}_N\otimes g(\bm{s},\bm{s}) & 0 \\ 
-\epsilon L\otimes \mathrm{D}h(0) & -\epsilon \mathbb{I}_N & 0 \\
-\epsilon Q^TC\otimes \mathrm{D}h(0) & 0 & -\epsilon \mathbb{I}_{N^2-N}
\end{pmatrix}.
\end{multline*}
These equations yield that there are $Nd+N$ coupled differential equations left
\begin{align}\label{eq:epsilonReduction}
\frac{\mathrm{d}}{\mathrm{d}t}
\begin{pmatrix}
\hat{{\bm{\xi}}} \\
\tilde{{\bm{\chi}}}
\end{pmatrix}
=
\begin{pmatrix}
S & -{\sigma} \mathbb{I}_N\otimes g(\bm{s},\bm{s}) \\
-\epsilon L\otimes \mathrm{D}h(0) & -\epsilon \mathbb{I}_N
\end{pmatrix}
\begin{pmatrix}
\hat{{\bm{\xi}}} \\
\tilde{{\bm{\chi}}}
\end{pmatrix}
\end{align}
with $\tilde{{\bm{\chi}}}=\hat{{\bm{\chi}}}_1$ that determine the stability for the synchronous state, and $N^2-N$ slave equations
\begin{align*}
\frac{\mathrm{d}}{\mathrm{d}t}
\bar{{\bm{\chi}}}
=
\begin{pmatrix}
-\epsilon Q^TC\otimes \mathrm{D}h(0) & 0 & -\epsilon \mathbb{I}_{N^2-N}
\end{pmatrix}
\begin{pmatrix}
\hat{{\bm{\xi}}} \\
\tilde{{\bm{\chi}}} \\
\bar{{\bm{\chi}}}
\end{pmatrix}
\end{align*}
with $\bar{{\bm{\chi}}}=(\hat{{\bm{\chi}}}_2^T,\dots,\hat{{\bm{\chi}}}_N^T)^T$ which are driven by the variables $\hat{{\bm{\xi}}}$ and $\tilde{{\bm{\chi}}}$ and, hence, can be solved explicitly once the latter once are known. By assumption, there is a unitary matrix $D_{L}=U^H L U $ where $D_{L}$ is the diagonalization of the Laplacian matrix $L$. Transforming the differential equation~\eqref{eq:epsilonReduction} by using the unitary transformation $U$, we get
\begin{widetext}
	\begin{align*}
	\frac{\mathrm{d}}{\mathrm{d}t}
	\begin{pmatrix}
	{{\bm{\zeta}}} \\
	{{\bm{\kappa}}}
	\end{pmatrix}
	=
	\begin{pmatrix}
	\mathbb{I}_{N}\otimes Df(\bm{s})+{\sigma} h(0)\left(r\mathbb{I}_N\otimes\mathrm{D_1}g(\bm{s},\bm{s}) + (r\mathbb{I}_N-D_L)\otimes \mathrm{D_2}g(\bm{s},\bm{s})\right) & -{\sigma} \mathbb{I}_N\otimes g(\bm{s},\bm{s}) \\
	-\epsilon D_{L}\otimes \mathrm{D}h(0) & -\epsilon \mathbb{I}_N
	\end{pmatrix}
	\begin{pmatrix}
	{{\bm{\zeta}}} \\
	{{\bm{\kappa}}}
	\end{pmatrix}
	\end{align*}
\end{widetext}
where $\begin{pmatrix}
U\otimes\mathbb{I}_d & 0\\
0 & U
\end{pmatrix}\begin{pmatrix}
\hat{{\bm{\xi}}} \\
\tilde{{\bm{\chi}}}
\end{pmatrix}=\begin{pmatrix}
{{\bm{\zeta}}} \\
{{\bm{\kappa}}}
\end{pmatrix}$.

Remarkably, the master stability function $\Lambda$ depends explicitly on the row sum $r$. Moreover, the master stability function seems to depend on $\sigma$, $r$, and $\mu$ independently. The time scale separation parameter $\epsilon$ is always kept fixed. However, in any case, one parameter can be disregarded. To see this, we note that the solution to the Eq.~\eqref{eq:suppl_adaptiveNW_MSF_kappa} is explicitly solvable and the solution reads
\begin{align*}
\kappa = \kappa_0 e^{-\epsilon(t-t_0)} - \epsilon \mu \mathrm{D}h(0) \int_{t_0}^{t} e^{-\epsilon(t-t')} \zeta(t')\,dt',
\end{align*}
where the first term vanishes for $t \to \infty$ and hence can be neglected (when studying asymptotic stability for $t\to\infty$). We use this and rewrite the asymptotic dynamics of~\eqref{eq:suppl_adaptiveNW_MSF_zeta}--\eqref{eq:suppl_adaptiveNW_MSF_kappa} in its integro-differential form
\begin{multline*}%\label{eq:MSF_FHN_int_diff}
\frac{\mathrm{d}\zeta}{\mathrm{d}t} = \left(\mathrm{D}f(\bm{s})+\sigma r h(0)\left(\mathrm{D_1}g(\bm{s},\bm{s})\right.\right. \\
\left.\left.\quad\quad + (1-\frac{\mu}{r}) \mathrm{D_2}g(\bm{s},\bm{s})\right)\right) \zeta\\
+ \epsilon \sigma r \frac{\mu}{r} g(\bm{s},\bm{s}) \mathrm{D}h(0) \int_{t_0}^{t} e^{-\epsilon(t-t')} \zeta(t')\,dt'.
\end{multline*}
Hence, the master stability function can be regarded as a function of two parameters, i.e., $\Lambda(\sigma,\mu,r)=\Lambda(\sigma r, \mu/r)$. Furthermore, in case of diffusive coupling, i.e., $g(\bm{x},\bm{y}) = g(\bm{x}-\bm{y})$, the master stability function can be regarded as a function of only one parameter $\Lambda(\sigma,\mu,r)=\Lambda(\sigma\mu)$.

%--------------------------------------------------
% The master stability function of adaptive phase oscillator networks
%--------------------------------------------------
\section{Master stability function for adaptive phase oscillator networks}
In this section, we provide a brief analysis of the master stability function for the adaptive Kuramoto-Sakaguchi network~\eqref{eq:APO_phi}--\eqref{eq:APO_kappa} of the main text. Using the result of Section~\ref{sec:Aug_MSA_MltPlx}, the stability of the synchronous state of system~\eqref{eq:APO_phi}--\eqref{eq:APO_kappa} of the main text is governed by the two differential equations
\begin{align*}
\frac{\mathrm{d}}{\mathrm{d}t} \begin{pmatrix}
\zeta \\
\kappa
\end{pmatrix}&= \begin{pmatrix}
\mu\sigma{\cos(\alpha)\sin(\beta)} &  - \sigma{\sin(\alpha)}\\
-\epsilon\mu \cos(\beta) & -\epsilon
\end{pmatrix}\begin{pmatrix}
\zeta \\
\kappa
\end{pmatrix},
\end{align*}
where $\mu\in \mathbb{C}$ stands for all eigenvalues of the Laplacian matrix $L$ corresponding to the base network described by the adjacency matrix $A$. The characteristic polynomial in $\lambda$ of the latter system is of degree two and reads
\begin{align}
\lambda^2 + \left(\epsilon-\sigma{\mu}\cos(\alpha)\sin(\beta)\right)\lambda -\epsilon\sigma{\mu}\sin(\alpha+\beta) = 0. \label{eq:charpoly}
\end{align}
The master stability function is given as $\Lambda(\sigma\mu)=\max(\mathrm{Re}(\lambda_1),\mathrm{Re}(\lambda_2))$ where $\lambda_1$ and $\lambda_2$ are the two solutions of the quadratic polynomial \eqref{eq:charpoly}. Figure~\ref{fig:MSF_Kuramoto} of the main text displays the master stability function for different parameters.

The boundary of the region in $\sigma{\mu}$ parameter space that corresponds to stable local dynamics, is given by $\lambda=\mathrm{i}\gamma$ with $\gamma \in \mathbb{R}$. Plugging this into Eq.~\eqref{eq:MSF_Kuramoto_eigenvalues} of the main text, we obtain
\begin{align*}
\sigma\mu = Z(\gamma)=a(\gamma)+\mathrm{i}b(\gamma) 
\end{align*}
with
\begin{align*}
a(\gamma) &= \epsilon\frac{\gamma^2\left(\cos\alpha\sin\beta-\sin(\alpha+\beta)\right)}{{\gamma}^2\cos^2\alpha\sin^2\beta+\epsilon^2\sin^2(\alpha+\beta)}, \\
b(\gamma) &=\frac{\gamma^3\cos\alpha\sin\beta+\epsilon^2\gamma\sin(\alpha+\beta)}{{\gamma}^2\cos^2\alpha\sin^2\beta+\epsilon^2\sin^2(\alpha+\beta)}. 
\end{align*}
Due to the symmetry of the master stability function, a necessary condition to observe a stability island is that the curve $\sigma\mu({\gamma})$ possesses two crossings with the real axis, i.e., two real solutions for $b(\gamma)=0$. The three crossings are given by $\gamma_1=0$ and as real solutions $\gamma_2$ and $\gamma_3$ of $\gamma^2\cos\alpha\sin\beta=-\epsilon^2\sin(\alpha+\beta)$. From this we deduce the existence condition for stability islands: $\sin(\alpha+\beta)/(\cos\alpha\sin\beta)<0$ ($\epsilon>0$). Note that $a(\gamma_2)=a(\gamma_3)$.

%--------------------------------------------------
% Measuring the amount of coherence in system of phase oscillators - the cluster parameter
%--------------------------------------------------
\section{The cluster parameter}
In this section, we introduce the cluster parameter $R_C$ as a measure for coherence in a system of coupled phase oscillators. A measure that can be used in order to detect frequency synchronization between two oscillators relies on the mean phase velocity (average frequency) of each phase oscillator
\begin{align}
\Omega_{i}=\lim_{T\to\infty}\frac{1}{T}\left(\phi_i(t_0+T)-\phi_i(t_0)\right).
\end{align} 
The frequency synchronization measure between nodes is given by
\begin{align}
\Omega_{ij} = \begin{cases}
1, \text{if }\Omega_i -\Omega_j=0,\\
0, \text{otherwise}.
\end{cases}
\end{align}
Numerically the limit is approximated by a very long averaging window. In addition, we use a sufficiently small threshold $\varpi$ in order to detect frequency synchronization numerically, i.e., $\Omega_{ij} = 1$ if $\Omega_i -\Omega_j<\varpi$. For the analysis presented here and in the main text, we use $\varpi=0.001$. Using the measure $\Omega_{ij}$, we define the cluster parameter
\begin{align}\label{eq:cluster_param}
R_C=\frac{1}{N^2}\sum_{i,j=1}^N\Omega_{ij}.
\end{align}
The cluster parameter measures the following. First, for each frequency cluster, the total number of pairwise synchronized nodes is computed. Second, all pairs of two nodes from the same cluster are summed up and normalized by the number of all possible pairs of nodes $N^2$. In case of full synchronization, frequency clustering, or incoherence the values of the cluster parameter are $R_C=1$, $1<R_C<0$, or $R_C=0$, respectively. A similar measure can be found in Refs.~\cite{KAS17,KAS18a}.

%--------------------------------------------------
% Desynchronization transition and the formation of partial synchronization patterns in adaptive phase oscillator networks
%--------------------------------------------------
\section{Desynchronization transition and the formation of partial synchronization patterns in adaptive phase oscillator networks}
In this section, we provide further details on the desynchronization transition in a network of adaptively coupled phase oscillators~\eqref{eq:APO_phi}--\eqref{eq:APO_kappa}.

\begin{figure}[h!]
	\includegraphics{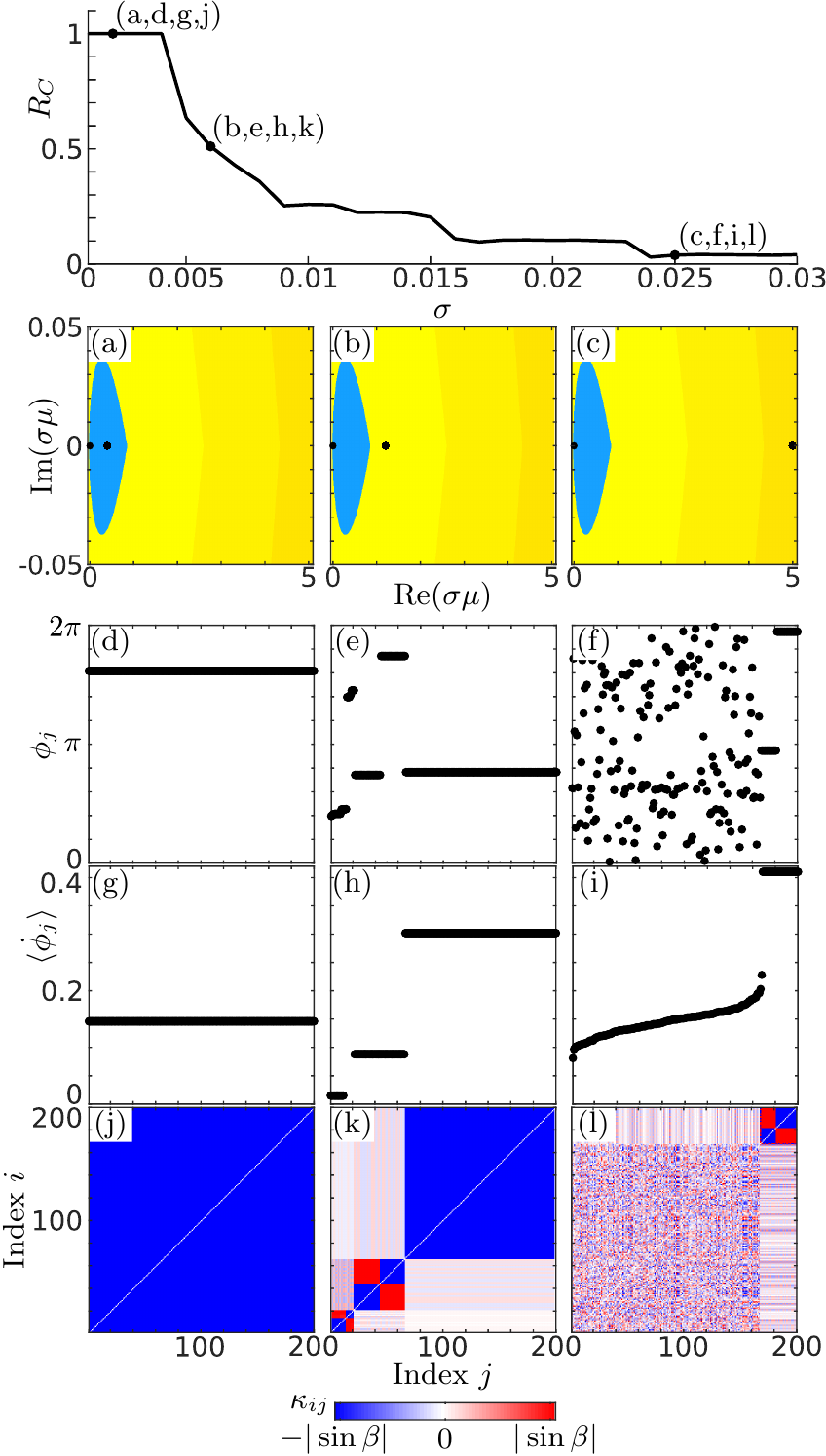}
	\caption{Dynamics in a globally coupled network of $200$ phase oscillators Eqs.~\eqref{eq:APO_phi}--\eqref{eq:APO_kappa} of the main text for different values of overall coupling strength $\sigma$ than in Fig.2 of the main text. Adiabatic continuation for increasing $\sigma$ with the stepsize of $0.001$, starting with the synchronous state $\phi_i=0$, $\kappa_{ij}=-a_{ij}\sin\beta$. The top panel shows the cluster parameter $R_C$ vs $\sigma$. For the three values of $\sigma$: (a,d,g,j) $\sigma=0.002$, (b,e,h,k) $\sigma=0.006$, and (c,f,i,l) $\sigma=0.025$,	the plots show: in (a,b,c) the master stability function (color coded as in Fig. 1 of the main text), together with $\sigma\mu_i$, where $\mu_i$ are the $N$ Laplacian eigenvalues of $A$; in (d,e,f) snapshots for $\phi_i$ at $t=30000$; in (g,h,i) the temporal average of the phase velocities $\langle\dot{\phi}_i\rangle$ over the last $5000$ time units; and in (j,k,l) snapshots for the coupling matrix $\kappa_{ij}$ at $t=30000$. Other parameters: $\alpha=0.49\pi$, $\beta=0.88\pi$, $\epsilon=0.01$. \label{fig:MSF_global_MC}}
\end{figure}
Figure~\ref{fig:MSF_global_MC} shows the cluster parameter $R_C$ for different values of the coupling constant $\sigma$. In the adiabatic continuation, we increase $\sigma$ step-wise after an integration time of $t=10000$. For each simulation, the final state of the previous simulations is taken as the initial condition with an additional small perturbation. Note that $R_C=1$ refers to full in-phase synchrony of the oscillators. We observe that, for small $\sigma$, the synchronous state is stable, see Fig.~\ref{fig:MSF_global_MC}(d,g,j). Here, the stability of the synchronous state is directly implied by the master stability function. We note that all Laplacian eigenvalues $\mu_i$ of a globally coupled network are given by either $\mu_i=0$ or $\mu_i=N$. In Figure~\ref{fig:MSF_global_MC}(a), all master function parameters $\sigma\mu$ lie within the stability island. 

By increasing the coupling constant, the values $\sigma\mu_i$ move out of the stability regions and the synchronous state becomes unstable. For intermediate values of $\sigma$ the emergence of multiclusters with hierarchical structure in the cluster size are observed. In Figure~\ref{fig:MSF_global_MC}(e,h,k) a multicluster states is shown with three clusters. Note that for the system~\eqref{eq:APO_phi}--\eqref{eq:APO_kappa} of the main text, in-phase synchronous and antipodal clusters have the same properties~\cite{BER19,BER19a}. In Refs.~\cite{BER19,BER19a} the role of the hierarchical structure of the cluster sizes have been discussed. Increasing the coupling constant further shows the emergence of incoherence. In Figure~\ref{fig:MSF_global_MC}(f,i,l), we show the coexistence of a coherent and an incoherent cluster. These states, also called chimera-like states, have been numerically analyzed in Refs.~\cite{KAS17,KAS18,KAS18a}.

\section{Network of coupled FitzHugh-Nagumo neurons with synaptic plasticity}
In this section, we describe the model of coupled FitzHugh-Nagumo neurons with synaptic plasticity and present the synchronous state used in the main text. The model is given by
\begin{align}
\tau \dot u_i &= u_i - \frac{u_i^3}{3} - v_i - {\sigma}\sum_{j = 1}^{N} a_{ij} 
\kappa_{ij} {u_i I_j},\label{eq:suppl_AFHN_FHN1}\\
\dot v_i &= u_i + a - b v_i,\label{eq:suppl_AFHN_FHN2}\\
\dot I_i &=  \alpha(u_i) (1-I_i)- {I_i}/{\tau_\text{syn}},\label{eq:suppl_AFHN_FHN3}\\
\dot{\kappa}_{ij} &= -\epsilon \left( \kappa_{ij} + a_{ij} e^{-\beta_1 (u_i - u_j + \beta_2)^2} \right),\label{eq:suppl_AFHN_kappa}
\end{align}
where $\alpha(u_i) = 2/(0.08 (1 + \exp(-{u_i}/0.05)))$, see Eqs.~\eqref{eq:AFHN_FHN1}--\eqref{eq:AFHN_kappa} of the main text. All variables and parameters are explained in Tab.~\ref{tab:AFHN_parameters}.
\begin{table}
	\begin{tabular}{p{0.3\columnwidth}p{0.65\columnwidth}}
		$u_i$ & membrane potential/activator\\
		$v_i$ & recovery/inhibitor variable\\
		$I_i$ & synaptic output variable \\
		$\kappa_{ij}$ & variable coupling weights\\
		$N$ & number of oscillators\\
		$a_{ij}$ & entries of adjacency matrix, $a_{ij}\in\{0,1\}$\\
		$\sigma$ & overall coupling strength\\
		$r$ & row sum, i.e., $r = \sum_{j=1}^{N}a_{ij}$\\
		$a=0.7, b=0.2$ & bifurcation parameters of the FitzHugh-Nagumo neuron\\
		$\tau=0.08$ & controls time separation between fast activation and slow inhibition\\
		$\epsilon=0.01$ & controls time separation between fast oscillation and slow adaptation\\
		$\tau_\text{syn}=5/6$ & synaptic decay rate\\
		$u_\text{shp}=0.05$ & coupling shape parameter\\
		$\beta_1,\beta_2$ & adaption control parameters		
	\end{tabular}
	\caption{\label{tab:AFHN_parameters}The table provides the meaning for each variable and parameter used in~\eqref{eq:suppl_AFHN_FHN1}--\eqref{eq:suppl_AFHN_kappa}.}
\end{table}
The form of the synaptic plasticity is similar to the rules used in~\cite{YUA11,CHA17a}. We introduce $\beta_1$ and $\beta_2$ as control parameters. In particular, we have $\beta_1=-h(0)/(2D h(0)\beta_2)$ and $\beta_2 =(2\mathrm{D} h(0)_{1}\ln( \mathrm{D} h(0)_{1}))/h(0)$ where $\mathrm{D} h(0)_{1}$ denotes the first component of $\mathrm{D} h(0)$.

The synchronous state of the equations~\eqref{eq:suppl_AFHN_FHN1}--\eqref{eq:suppl_AFHN_kappa} is given by a solution of
\begin{align}
\tau \dot u_s &= u_s - \frac{u_i^3}{3} - v_s + {\sigma r}{u_s I_s}e^{-\beta_1 \beta_2^2},\label{eq:AFHN_FHN1_sync}\\
\dot v_s &= u_s + a - b v_s,\label{eq:AFHN_FHN2_sync}\\
\dot I_s &=  \alpha(u_s) (1-I_s)- {I_s}/{\tau_\text{syn}},\label{eq:AFHN_FHN3_sync}\\
{\kappa}^s_{ij} &= - a_{ij} e^{-\beta_1 \beta_2^2},\label{eq:AFHN_kappa_sync}
\end{align}
where $(u_i,v_i,I_i) = \bm{s} = (u_s,v_s,I_s)$ for all $i=1,\dots,N$. 
\begin{figure}
	\includegraphics{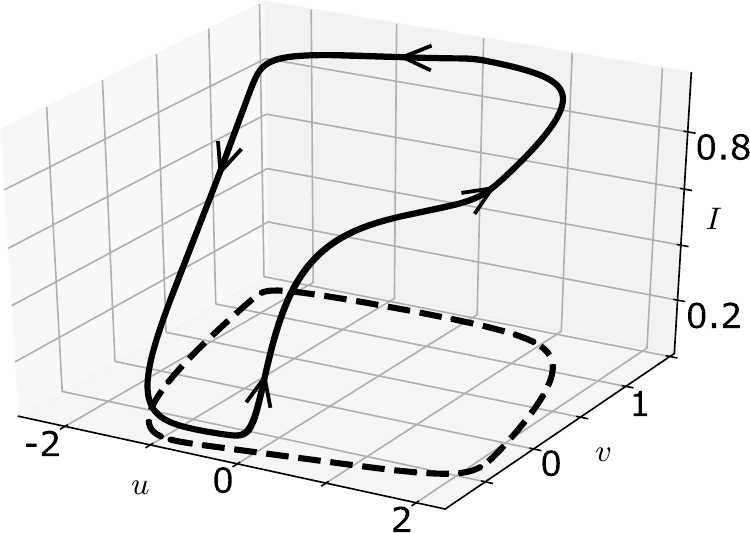}
	\caption{\label{fig:AFHN_sync} Limit cycle in Eqs.~\eqref{eq:AFHN_FHN1_sync}--\eqref{eq:AFHN_FHN3_sync} as solid line and the projection onto the $u$-$v$-plane as dashed line. Parameters: $\sigma=0.002$, $r=200$, $h(0)=0.8$ and $\mathrm{D}h(0)=(80,0,0)$. All other parameters as in Tab.~\ref{tab:AFHN_parameters}.}
\end{figure}
In Fig.~\ref{fig:AFHN_sync}, we display a limit cycle as a stable numerical solution of~\eqref{eq:AFHN_FHN1_sync}--\eqref{eq:AFHN_kappa_sync} for the set of parameters used in the main text.% In fact, we obtain a periodic orbit as the synchronous state.

%--------------------------------------------------
% The master stability function and desynchronization transition in adaptive networks of FitzHugh-Nagumo neurons
%--------------------------------------------------
\section{The master stability function and desynchronization transition in adaptive networks of FitzHugh-Nagumo neurons}
In this section, we consider the model of adaptively coupled FitzHugh-Nagumo neurons~\eqref{eq:suppl_AFHN_FHN1}--\eqref{eq:suppl_AFHN_kappa}. We give insights into the derivation of the system's master stability function as well as on the desynchronization transition induced by the adaptivity. 

In order to investigate the local stability of the synchronous states that solves Eqs.~\eqref{eq:AFHN_FHN1_sync}--\eqref{eq:AFHN_kappa_sync}, see Fig.~\ref{fig:AFHN_sync}, we linearize Eqs.~\eqref{eq:suppl_AFHN_FHN1}--\eqref{eq:suppl_AFHN_kappa} around these states. Using the results of Section~\ref{sec:Aug_MSA_MltPlx}, the stability of the synchronous solution is governed by the set of equations
\begin{align*}
\begin{split}
\frac{\mathrm{d}\zeta}{\mathrm{d}t} &= \bigg(\mathrm{D}f(\bm{s})+\sigma r h(0)\big(\mathrm{D_1}g(\bm{s},\bm{s}) \\
&\quad\quad + (1-\frac{\mu}{r}) \mathrm{D_2}g(\bm{s},\bm{s})\big)\bigg) \zeta - {\sigma} g(\bm{s},\bm{s}) \kappa, 
\end{split}\\ 
\frac{\mathrm{d}\kappa}{\mathrm{d}t} &=  -\epsilon\left(\mu \mathrm{D}h(0) \zeta + \kappa\right).
\end{align*}
Here, the derivatives of the functions $f$, $g$, and $h$ are
\begin{align*}
{\mathrm{D}f}(\bm s) &= \begin{pmatrix}
\frac{1}{\tau}\left(1-u_s^2\right) & -\frac{1}{\tau} & 0 \\
1 & -b & 0 \\
\frac{\tau(\alpha(u_s))^2 \left(1-I_s\right)}{\alpha_0 u_{\text{shp}} \exp(\frac{u_s}{u_{\text{shp}}})} & 0 & -\alpha(u_s) - \frac{1}{\tau_\text{syn}} \\
\end{pmatrix},\\
{\mathrm{D_1}g}(\bm s,\bm s) &= \begin{pmatrix}
I_s & 0 & 0 \\
0 & 0 & 0 \\
0 & 0 & 0 \\
\end{pmatrix},\\
{\mathrm{D_2}g}(\bm s,\bm s) &= \begin{pmatrix}
0 & 0 & u_s \\
0 & 0 & 0 \\
0 & 0 & 0 \\
\end{pmatrix},\\
{\mathrm{D}h}(0) &= \begin{pmatrix}
-2\beta_1\beta_2\exp(-\beta_1\beta_2^2) & 0 & 0 \\
\end{pmatrix}.
\end{align*}

Using this, we are able to determine numerically the maximum Lyapunov exponents and hence the stability of the periodic orbit displayed in Fig.~\ref{fig:AFHN_sync}.
\begin{figure}
	\includegraphics{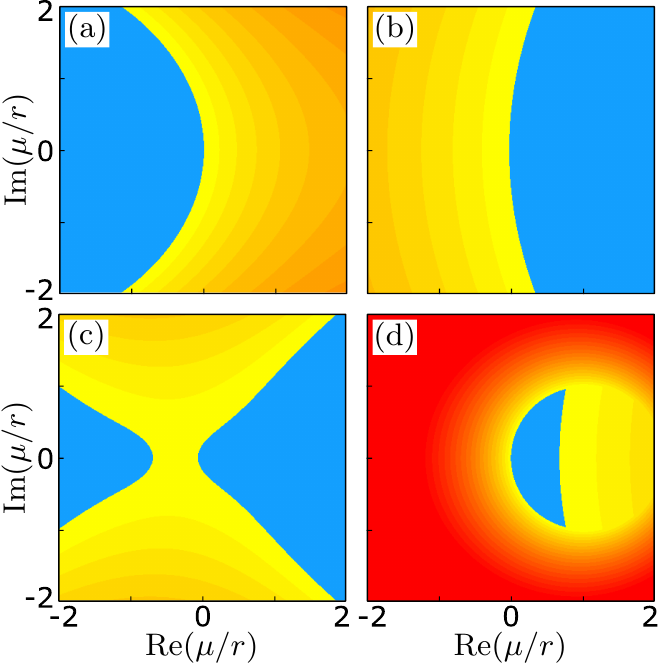}
	\caption{The master stability functions for the synchronous solution of~\eqref{eq:suppl_AFHN_FHN1}--\eqref{eq:suppl_AFHN_kappa} and different plasticity rules are displayed (color code as in Fig.~\ref{fig:MSF_Kuramoto} of the main text). Regions belonging to negative Lyapunov exponents are colored blue. Parameters: the control parameters $\beta_1$ and $\beta_2$ are chosen such that (a) $h(0)=0.8$, $\mathrm{D}h(0)=(50,0,0)$ (b) $h(0)=-0.2$, $\mathrm{D}h(0)=(0,0,0)$, (c) $h(0)=0.8$, $\mathrm{D}h(0)=(10,0,0)$, and (d) $h(0)=0.4$, $\mathrm{D}h(0)=(50,0,0)$. The overall coupling constant is set to $\sigma=0.005$. All other parameters are as in Fig.~\ref{fig:AFHN_sync}. \label{fig:Suppl_MSF_FHN}}
\end{figure}
In Fig.~\ref{fig:Suppl_MSF_FHN}, we show different shapes of the master stability function depending on the form of the plasticity rule, i.e., depending on $h(0)$ and $D h(0)$. We observe that for certain parameters almost complete half spaces in the $\mu/r$-plane refer to stable or unstable local dynamics, see Fig.~\ref{fig:Suppl_MSF_FHN}(a,b). This is similar to Fig.~\ref{fig:MSF_Kuramoto}(d,e) of the main text where we display the master stability function of the phase oscillator model. Most remarkably, similar to the phase oscillator model~\eqref{eq:APO_phi}--\eqref{eq:APO_kappa} we find parameters for which stability islands exist, see Fig.~\ref{fig:Suppl_MSF_FHN}(d).

\begin{figure}
	\includegraphics{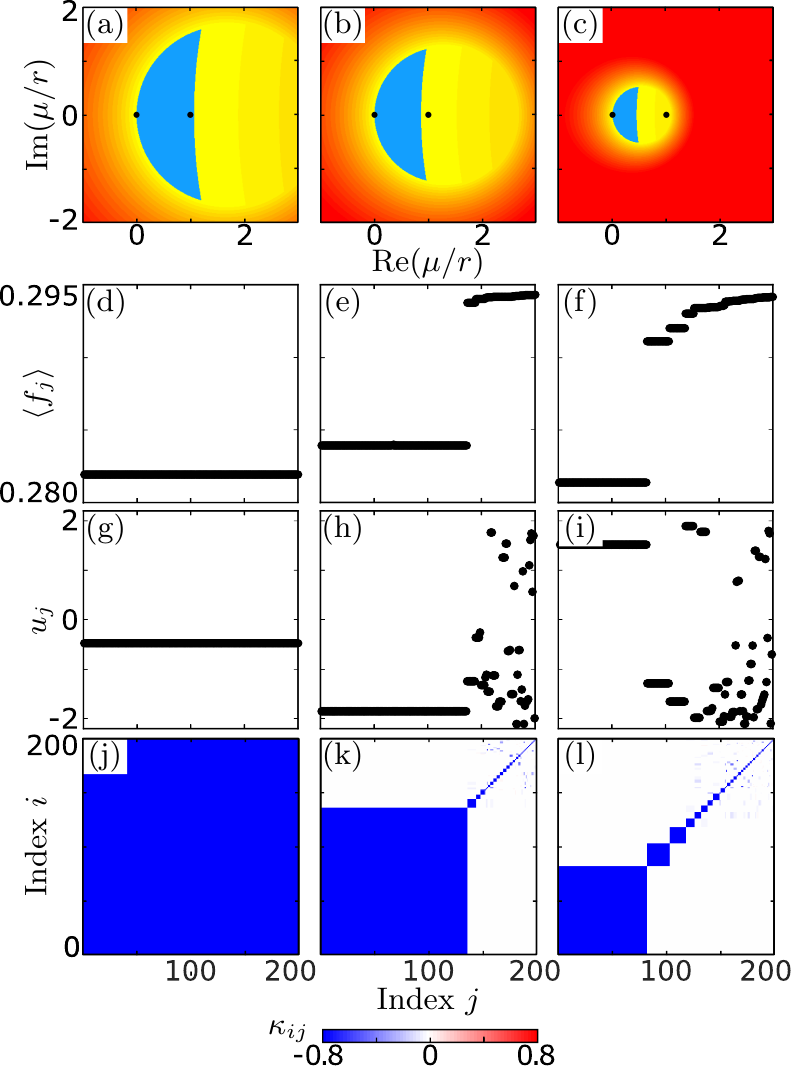}
	\caption{Dynamics of globally coupled network of $200$ FitzHugh-Nagumo neurons with plasticity Eqs.~\eqref{eq:AFHN_FHN1}--\eqref{eq:AFHN_kappa}. Adiabatic continuation for an increasing overall coupling strength $\sigma$ with the step size $0.0005$, starting with the synchronous state of Eqs.~\eqref{eq:AFHN_FHN1}--\eqref{eq:AFHN_kappa}. For the three values of $\sigma$: (a,d,g,j) $\sigma=0.002$, (b,e,h,k) $\sigma=0.0025$, and (c,f,i,l) $\sigma=0.005$, the plots show: in (a,b,c), the master stability function, together with $\mu_i/r$, where $\mu_i$ are the $N$ Laplacian eigenvalues (color code as in Fig.~\ref{fig:MSF_Kuramoto} of the main text), in (d,e,f) the average frequency $\langle f_i\rangle$, in (g,h,i) snapshots for $u_i$ at $t=10000$, and in (j,k,l) snapshots for the coupling matrices $\kappa_{ij}$ at $t=10000$. Here $\langle f_i\rangle=M_i/1000$, where  $M_i$ is the number of rotations (spikes) of neuron $i$ during the time interval of length 1000. The control parameters for the adaptation rule $\beta_1$ and $\beta_2$ are chosen such that $h(0)=0.8$ and $\mathrm{D}h(0)=(80,0,0)$. All other parameters can be taken from Tab.~\ref{tab:AFHN_parameters}.\label{fig:Suppl_MSF_global_FHN}}
\end{figure}
As we know from the example of phase oscillators, the presence of a stability island may induce a desynchronization transition for an increasing overall coupling strength $\sigma$. In order to show this transition, we follow the same approach already presented in~Fig.~\ref{fig:MSF_global_MC}. The results of the adiabatic continuation on a globally coupled network are shown in Fig.~\ref{fig:Suppl_MSF_global_FHN}. We note that in contrast to the case of phase oscillators, here, the shape of the master stability function depends explicitly on $\sigma$. The desynchronization is described in the main text. Additionally to the figure given in the main text, we provide plots for the coupling matrices in Fig.~\ref{fig:Suppl_MSF_global_FHN}(j,k,l). The coupling matrices show very nicely the emergence of partial synchronization structures in the transition from coherence to incoherence which is induced by the stability island.

%--------------------------------------------------
% \section{Example of a complex network structure}
%--------------------------------------------------
%\section{Example of a complex network structure}
\begin{figure}[h!]
	\includegraphics{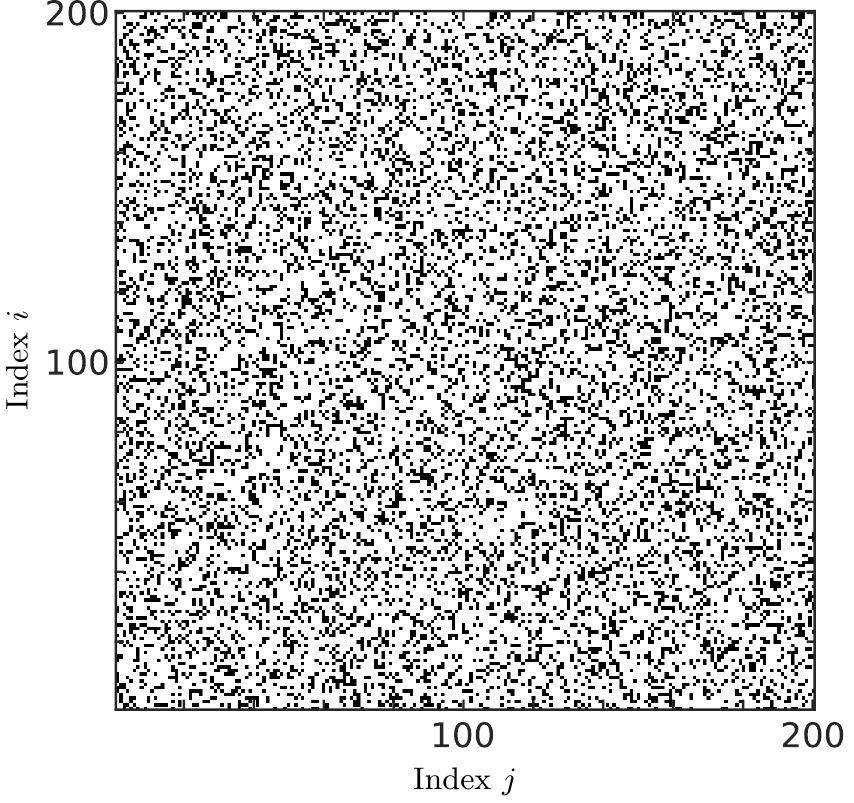}
	\caption{Adjacency matrix $A_c$ of a connected, directed random network of $N=200$ nodes with constant row sum $r=50$. The illustration shows the adjacency matrix where black and white refer to whether a link between two nodes exist or not, respectively.\label{fig:Suppl_RandomMat}}
\end{figure}
%The adjacency matrix $A_c$ displayed in Fig.~\ref{fig:Suppl_RandomMat} is obtained by the following procedure. For each node $i$ of the $N$ nodes, $r$ links are randomly (with uniform distribution) picked from the set consisting of all links from a node $j\ne i$ to node $i$. This procedure results in a random directed network with $N$ nodes and constant row sum (in-degree) $r$. After the procedure we test if the resulting network is connected.
%merlin.mbs apsrev4-1.bst 2010-07-25 4.21a (PWD, AO, DPC) hacked
%Control: key (0)
%Control: author (8) initials jnrlst
%Control: editor formatted (1) identically to author
%Control: production of article title (-1) disabled
%Control: page (0) single
%Control: year (1) truncated
%Control: production of eprint (0) enabled
%

\end{document}